\documentclass[9pt,twocolumn,twoside]{opticajnl}
\journal{opticajournal} 

\setboolean{shortarticle}{false}

\usepackage{lineno}

\title{Optical systems with translational invariance}

\author
{Dmitry Zhuridov}

\affil
{Institute of Theoretical Physics, University of Wroc\l aw, Plac Maxa Borna 9, 50204, Wroc\l aw, Poland\\
dmitry.zhuridov@uwr.edu.pl}

\begin{abstract}
	The leading order monochromatic aberrations are investigated for the optical systems, which obey single-plane symmetry and translational invariance. These aberrations are classified from the symmetry principles for the wave aberration function. Fermat's Principle is applied for a system made of generic cylindrical surfaces and a complete set of the aberration coefficients required for calculation of the aberrations for any paraxial ray is obtained. To demonstrate the applied value of the analytical results obtained the criteria of compensation of the leading aberrations for cylindrical analog of the Cassegrain telescope were found.
\end{abstract}

\setboolean{displaycopyright}{false} 

\begin{document}
	
	\maketitle

\section{Introduction}
Rotational invariance limits the capabilities of optical systems (OS). Going beyond rotational symmetry allows one to expand these capabilities that plays increasing role for modern astronomical instrumentation. Recently, anamorphic optics, which is based on double-plane symmetry~\cite{Yuan-Sasian_I,Yuan-Sasian_II,Caron_Baumer_2020}, is intensively developing. In particular, pupil anamorphism was discussed in the context of polarimetry and high-resolution spectrography~\cite{Hanany,CORE:2017ssf}, wide field-of-view OS with anamorphic optical surfaces were investigating~\cite{Peschel:2014,Kashima:2017yub,Shi_Zheng}, and toroidal curved detectors were proposed for improving the performance of imaging~\cite{Muslimov}. 

Calculation of anamorphic OS due to its complexity is usually performed numerically, using a specific optical design software~\cite{YCLYG2017,Reshidko_Sasian_2018,Muslimov,Caron_Baumer_2020}. In this paper a class of relatively simple OS without axial symmetry is considered analytically in the spirit of classical textbooks, e.g.,~\cite{Born_Wolf,Slusarev,Schroeder}. We discuss a subclass of OS with single-plane symmetry~\cite{Stone_Howard_2016,Reshidko_Sasian_2018}, namely OS with translational invariance along a transverse direction. However, these OS can serve as components of more advanced anamorphic OS~\cite{Yuan-Sasian_II}. 

A generic approach to the leading-order aberrations of the OS with the translational invariance is discussed in the next section. Derivation of the transverse aberrations for a single cylindrical surface based on the Fermat's Principle is presented in section~\ref{sec:single_surface}. Finally, in section~\ref{sec:multisurface_systems} we discuss calculation of aberrations for multisurface OS with the translational invariance and demonstrate the power of presented analytical method on an example of cylindrical analog of the Cassegrain telescope, and then we conclude.

\section{Aberrations for cylindrical surfaces: generic picture}\label{sec:cyl}

The number of aberrations for generic optical surfaces is high. The wave aberration $\Phi$, which describes the optical path difference between the actual wavefront and the intended ideal one, is a function of four parameters~\cite{Born_Wolf}, e.g., the coordinates \{$x$, $y$, $x^\prime$, $y^\prime$\}  
of the cross-points $P_0(x,y,0)$ and $P^\prime(x^\prime,y^\prime,z^\prime)$ 
for a paraxial ray 
with the object plane and the exit pupil plane, respectively. These two planes are orthogonal to the $Z$ axis (in the case of rotatoinally symmetric OS it is the symmetry axis, while in the case of discussed translationally invariant OS it can be any axis that lies in the symmetry plane and is orthogonal to the vector of translation). 
Expansion of $\Phi$ into the four argument power series gives 35 terms in the fourth-order part $\Phi^{(4)}$, which derivatives determine the number of the third-order aberrations (see Eqs.~(\ref{eq:aberXdef}) and (\ref{eq:aberYdef}) below). 
In general, there are 20 different third-order power products for the four variables. 
Number of aberrations can be reduced for OS with definite symmetries, in particular, single-plane or double-plane symmetry. In the paraxial region a generic anamorphic OS (double-plane symmetric one) can be approximated 
by two independent rotationally symmetric OS, each associated with one symmetry plane~\cite{Yuan-Sasian_I}. This helps to derive 16 monochromatic primary aberration coefficients for anamorphic systems made from cylindrical surfaces~\cite{Yuan-Sasian_II}.  

The symmetry of a cylindrical surface, namely, single-plane symmetry plus translational invariance, 
allows for a simplified generic consideration and explicit analytical description presented in the following. Consider a system of cylindrical 
surfaces, which is symmetric under reflection in the $YZ$ plane and translationally invariant along the $Y$ axis. The translational invariance leaves three independent variables, e.g., $x$, $x^\prime$ and the difference $y-y^\prime$. We may construct from them the following six invariants of the second order
\begin{eqnarray}
x^2, \quad x^{\prime2}, \quad (y-y^\prime)^2, \quad xx^\prime, \quad x(y-y^\prime), \quad x^\prime(y-y^\prime).
\end{eqnarray}
The discussed plane symmetry leaves only four invariants of
\begin{eqnarray}\label{eq:ABCD}
A = x^2, \quad B = x^{\prime2}, \quad C = (y-y^\prime)^2, \quad D = xx^\prime,
\end{eqnarray}
which completely determine the ray position. Since the variables enter only in the quadratic combinations of \eqref{eq:ABCD}, it follows that $\Phi^{(4)}$ must be of the form 
\begin{eqnarray}\label{eq:W4}
&&\Phi^{(4)} = 
a_1A^2 + \frac{a_2}{4}B^2 + \frac{a_3}{4}C^2 + \frac{a_4}{2}D^2 \nonumber\\ 
&&+ \frac{a_5}{2}AB + \frac{a_6}{2}AC + a_7AD + \frac{a_8}{2}BC + a_9BD + a_{10}CD,
\end{eqnarray}
where $a_i$ are the constants, and the numerical factors of $1/2$ and $1/4$ are introduced for the further convenience.

In the coordinates $X^\prime Y^\prime Z^\prime$ associated with the exit pupil (specifically, $X$-$X^\prime$, $Y$-$Y^\prime$ and $Z$-$Z^\prime$ are the pairs of co-directed axes, and $X^\prime Y^\prime$ is the exit pupil plane), an equation of the Gaussian reference sphere, which is centered on the 
Gaussian image point $P_1^*(x_1^*,y_1^*,D_1)$ and 
passes through the vertex 
$O^\prime(0,0,0)$ of the exit pupil, can be written as
\begin{eqnarray}
(x^\prime-x_1^*)^2 + (y^\prime-y_1^*)^2 + (z^\prime-D_1)^2 = R^{\prime2},
\end{eqnarray}
where the refractive index of the material behind the exit pupil is considered equal to one for brevity, $O^\prime P_1^* = R^{\prime}$ is the radius of this sphere and $D_1$ is the distance between the exit pupil plane and the image plane. Following the consideration of Ref.~\cite{Slusarev}, we take into account the fact that for the case of small aberrations in the paraxial region the directions of normal to the Gaussian reference sphere and normal to the real wave surface for the respective actual ray $P^\prime P_1$ are about the same. Then equation for the real wave surface can approximately be written as $f(x^\prime,y^\prime,z^\prime)=0$ with
\begin{eqnarray}\label{eq:real_wave}
f
=(x^\prime-x_1^*)^2 + (y^\prime-y_1^*)^2 + (z^\prime-D_1)^2 - (s^{\prime}+\Phi)^2,
\end{eqnarray}
where $O^\prime P_1 = s^\prime$ is the distance from $O^\prime$ to the actual image point $P_1$, and $\Phi \ll R^\prime \approx s^\prime$ with the unity refractive index for short. Equation of the normal to the real wave in the point with the coordinates $x^\prime$, $y^\prime$ and $z^\prime$ can be expressed as
\begin{eqnarray}
\frac{X^\prime-x^\prime}{\frac{\partial f}{\partial x^\prime}} = \frac{Y^\prime-y^\prime}{\frac{\partial f}{\partial y^\prime}} = \frac{Z^\prime-z^\prime}{\frac{\partial f}{\partial z^\prime}}.
\end{eqnarray}
Using \eqref{eq:real_wave}, we have
\begin{eqnarray}\label{eq:normal}
\frac{X^\prime-x^\prime}{x^\prime-x_1^*-s^\prime\frac{\partial\Phi}{\partial x^\prime}} = \frac{Y^\prime-y^\prime}{y^\prime-y_1^*-s^\prime\frac{\partial\Phi}{\partial y^\prime}} = \frac{Z^\prime-z^\prime}{z^\prime - D_1}.
\end{eqnarray}
Substituting the coordinates of $P_1(x_1,y_1,D_1)$ into \eqref{eq:normal}, we get the following expressions for the components of transverse aberration 
\begin{eqnarray}
\delta_{1x} &\equiv& x_1 - x_1^* =  D_1 \frac{\partial \Phi}{\partial x^\prime}, \label{eq:aberXdef}\\
\delta_{1y} &\equiv& y_1 - y_1^* =  D_1 \frac{\partial \Phi}{\partial y^\prime}.\label{eq:aberYdef}
\end{eqnarray}
This result also comes from the difference between the slopes of the ray for a perfect system and the actual ray (Chapter~5 in Ref.~\cite{Schroeder}). 
Setting $y$ to zero in Eqs.~(\ref{eq:ABCD}) and (\ref{eq:W4}) by choosing $XZ$ plane, in the leading order we obtain
\begin{eqnarray}
\frac{\delta_{1x}}{D_1} &=&  (a_2B + a_5A + a_8C + 2a_9D)x^\prime \nonumber\\
&+& (a_4D + a_7A + a_9B + a_{10}C)x, \\
\frac{\delta_{1y}}{D_1} &=&  (a_3C + a_6A + a_8B + 2a_{10}D)y^\prime,
\end{eqnarray}
which in terms of $\tilde a_i \equiv D_1 a_i$ can be rewritten as
\begin{eqnarray}
\delta_{1x} &=&  (\tilde a_2x^{\prime2} + \tilde a_8y^{\prime2})x^\prime + (3\tilde a_9x^{\prime2} + \tilde a_{10}y^{\prime2})x \nonumber\\
&+& (\tilde a_4 + \tilde a_5)x^2x^\prime \quad+ \tilde a_7x^3, \label{eq:aberrations1}\\
\delta_{1y} &=&  (\tilde a_8x^{\prime2} + \tilde a_3y^{\prime2})y^\prime + 2\tilde a_{10}x x^\prime y^\prime + \tilde a_6x^2y^\prime,
\end{eqnarray}
where in the right-hand sides the first, second and third terms account for the three-component generalized spherical aberration that we call elliptic aberration for short ($e$), two-component modified coma ($c$) and astigmatism with 
the field curvature 
($a$), respectively, while the last term in \eqref{eq:aberrations1} represents the distortion ($d$). These aberrations are discussed one by one in the following. In total they contain 8 degrees of freedom since $\tilde a_4$ and $\tilde a_5$ parameters enter the same term. In case of $\tilde a_2=\tilde a_3=\tilde a_8$ and $\tilde a_9=\tilde a_{10}$ we end up with the standard set of five aberrations for the spherically symmetric OS.

\subsection{Elliptic aberrations}

The 
elliptic aberrations generalize the spherical one and depend on the three parameters $\tilde a_2$, $\tilde a_3$ and $\tilde a_8$. 
In the polar coordinates 
\begin{eqnarray}
x^\prime&=&\rho^\prime\cos\theta^\prime, \label{eq:polar1}\\
y^\prime&=&\rho^\prime\sin\theta^\prime  \label{eq:polar2}
\end{eqnarray}
the respective contribution to the transverse aberrations takes the form
\begin{eqnarray}
\delta_{1x}^e &=&  \rho^{\prime3}\left(\frac{\tilde a_2 + \tilde a_8}{2} + \frac{\tilde a_2 - \tilde a_8}{2}\cos2\theta^\prime\right) \cos\theta^\prime, \label{eq:ellip_ab1}\\
\delta_{1y}^e &=&  \rho^{\prime3}\left(\frac{\tilde a_3 + \tilde a_8}{2} + \frac{\tilde a_8 - \tilde a_3}{2}\cos2\theta^\prime\right) \sin\theta^\prime, \label{eq:ellip_ab2}
\end{eqnarray}
Using the following trigonometric identities for the products of cosines and a sine
\begin{eqnarray}
\cos\alpha\cos\beta &=& \frac{1}{2} \left[ \cos(\alpha+\beta) + \cos(\alpha-\beta) \right], \\
\cos\alpha\sin\beta &=& \frac{1}{2} \left[ \sin(\alpha+\beta) - \sin(\alpha-\beta) \right],
\end{eqnarray}
one can split Eqs.~(\ref{eq:ellip_ab1})--(\ref{eq:ellip_ab2}) into the two sets of parametric equations for the two ellipses:
\begin{eqnarray}
\delta_{1x}^{e_i} &=& a_{e_i}\cos\theta_{e_i},\\
\delta_{1y}^{e_i} &=& b_{e_i}\sin\theta_{e_i},
\end{eqnarray}
where $\theta_{e_1} = \theta^\prime$ with the semi-major and semi-minor axes defined by
\begin{eqnarray}
a_{e_1} &=& \frac{3\tilde a_2+\tilde a_8}{4}\rho^{\prime3}, \\
b_{e_1} &=& \frac{3\tilde a_3+\tilde a_8}{4}\rho^{\prime3}
\end{eqnarray}
for the first ellipse, and $\theta_{e_2} = 3\theta^\prime$ with the following semi-axes
\begin{eqnarray}
a_{e_2} &=& \frac{\tilde a_2-\tilde a_8}{4}\rho^{\prime3}, \\
b_{e_2} &=& \frac{\tilde a_8-\tilde a_3}{4}\rho^{\prime3}
\end{eqnarray}
for the second one.

\begin{figure}[ht!]
	\centering\includegraphics[width=4.2cm]{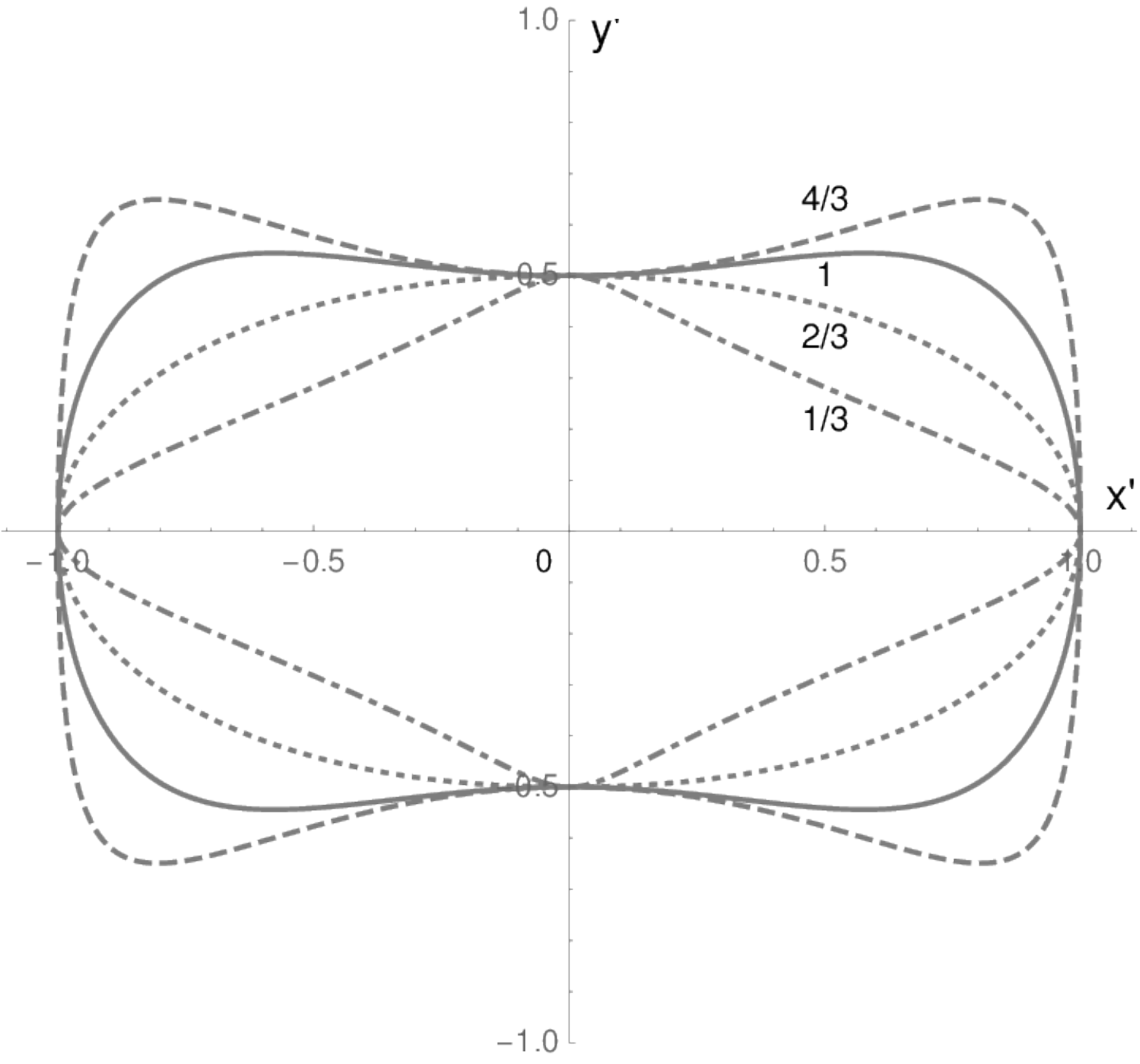}~~~~~
	\centering\includegraphics[width=4.2cm]{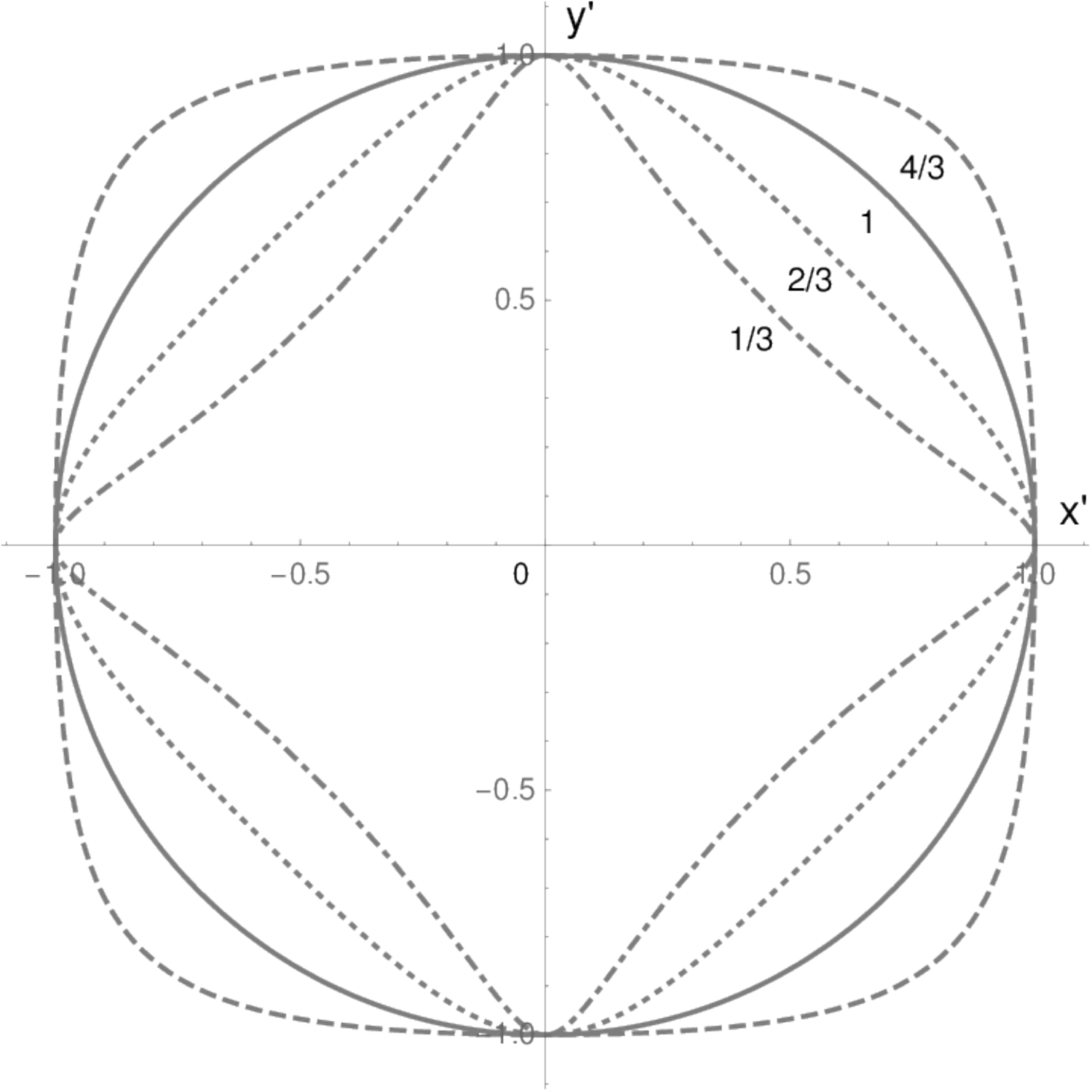}
	\caption{Elliptic aberrations in the 
		exit pupil plane $x^\prime y^\prime$ for \mbox{$\rho^\prime=\tilde a_2=1$}, $\tilde a_3=0.5$ ({\it left}) or $\tilde a_3=1$ ({\it right}), and the denoted values of $\tilde a_8$: 1/3, 2/3, 1 and 4/3 (in unspecified units).}\label{fig:0}
\end{figure}

In the case of $\tilde a_2=\tilde a_3$ these two ellipses 
degenerate to a superposition of the two parametric circles (with the opposite circle directions), 
which in turn degenerate to one circle in the case of \mbox{$\tilde a_2=\tilde a_3=\tilde a_8$} as shown in Fig.~\ref{fig:0}. This circle corresponds to a usual spherical aberration.

\subsection{Modified coma}

In the same coordinates defined by Eqs.~(\ref{eq:polar1}) and (\ref{eq:polar2}) the terms representing modified coma can be written~as
\begin{eqnarray}
\delta_{1x}^c - x_c  &=&  a_c \cos2\theta^\prime, \\
\delta_{1y}^c  &=&  b_c \sin2\theta^\prime,
\end{eqnarray}
where 
\begin{eqnarray}
x_c &=& \frac{3\tilde a_9 + \tilde a_{10}}{2}x\rho^{\prime2}, \\
a_c &=& \frac{3\tilde a_9 - \tilde a_{10}}{2}x\rho^{\prime2}, \\
b_c &=& \tilde a_{10}x\rho^{\prime2}.
\end{eqnarray}
This represents the ellipse shifted along the $X$ axis, which equation can be rewritten as
\begin{eqnarray}
\frac{(\delta_{1x}^c-x_c)^2}{a_c^2} + \frac{\delta_{1y}^{c2}}{b_c^2} = 1.
\end{eqnarray}
In the case of $\tilde a_9=\tilde a_{10}$ this aberration degenerates to usual coma.

\subsection{Astigmatism and field curvature}

In the same polar coordinates the respective terms can be written as
\begin{eqnarray}
\delta_{1x}^a  &=&  a_a \cos\theta^\prime, \\
\delta_{1y}^a  &=&  b_a \sin\theta^\prime,
\end{eqnarray}
where 
\begin{eqnarray}
a_a &=& (\tilde a_4 + \tilde a_5)x^2\rho^{\prime}, \\
b_a &=& \tilde a_6x^2\rho^{\prime}.
\end{eqnarray}
The above equations describe the ellipse (similarly to the case of spherically symmetric systems) and can be rewritten as
\begin{eqnarray}
\frac{\delta_{1x}^{a2}}{a_a^2} + \frac{\delta_{1y}^{a2}}{b_a^2} = 1.
\end{eqnarray}

\subsection{Distortion}

The 
distortion is determined by a single coordinate $x$ as
\begin{eqnarray}
\delta_{1x}^d  =  \tilde a_7 x^3.
\end{eqnarray}
It does not depend on the coordinates $x^\prime$ and $y^\prime$ of the point, in which the ray crosses the exit pupil. Similarly to the case of spherically symmetric OS, departure of the image point from the ideal image one due to the distortion is proportional to $x^3$.

\section{Aberrations for a single cylindrical surface}\label{sec:single_surface}
\subsection{Application of Fermat's Principle}

Consider a cylindrical optical surface that is invariant under translations along the $Y$ axis and given by the following equation
\begin{eqnarray}\label{eq:surface}
z = \frac{x^2}{2R} + \frac{x^4}{8} \left( \frac{1+\kappa}{R^3} + \frac{b}{n_1-n_0} \right) \equiv 
\frac{x^2}{2R} + \frac{\alpha x^4}{8}.
\end{eqnarray}
The homogeneous medium 
before the surface along the $Z$ axis has refractive index $n_0$ and the one 
after the surface has index $n_1$. 
We remark that \eqref{eq:surface} corresponds to the $XZ$ section of a generalized surface of rotation discussed in Ref.~\cite{Schroeder}, 
where $\kappa$ is the conic constant and the $b$ term explicitly includes the type of aspheric term required for corrector plates. In particular, $\alpha=0$ for parabolic cylinder, $\alpha=R^{-3}$ for cylinder of revolution and $\alpha=(1+\kappa)R^{-3}$ with $\kappa<-1$ for hyperbolic cylinder.

\begin{figure}[ht!]
	\centering\includegraphics[width=8.6cm]{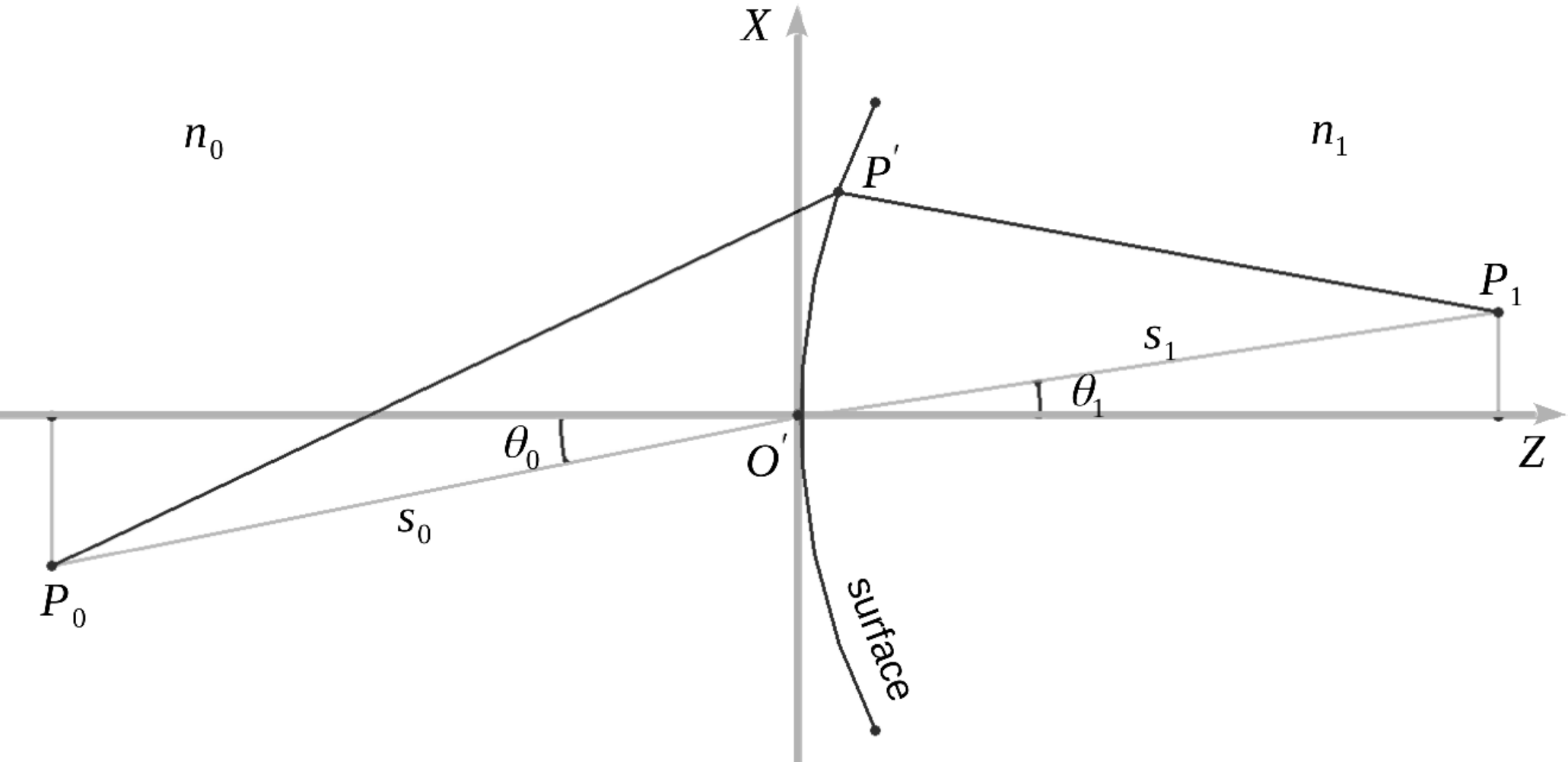}
	\caption{$XZ$ projection of the path of arbitrary ray through the refracting surface. The chief ray crosses the $Y$ axis in the point $O^\prime$.}\label{fig:1}
\end{figure}
\begin{figure}[ht!]
	\centering\includegraphics[width=8.6cm]{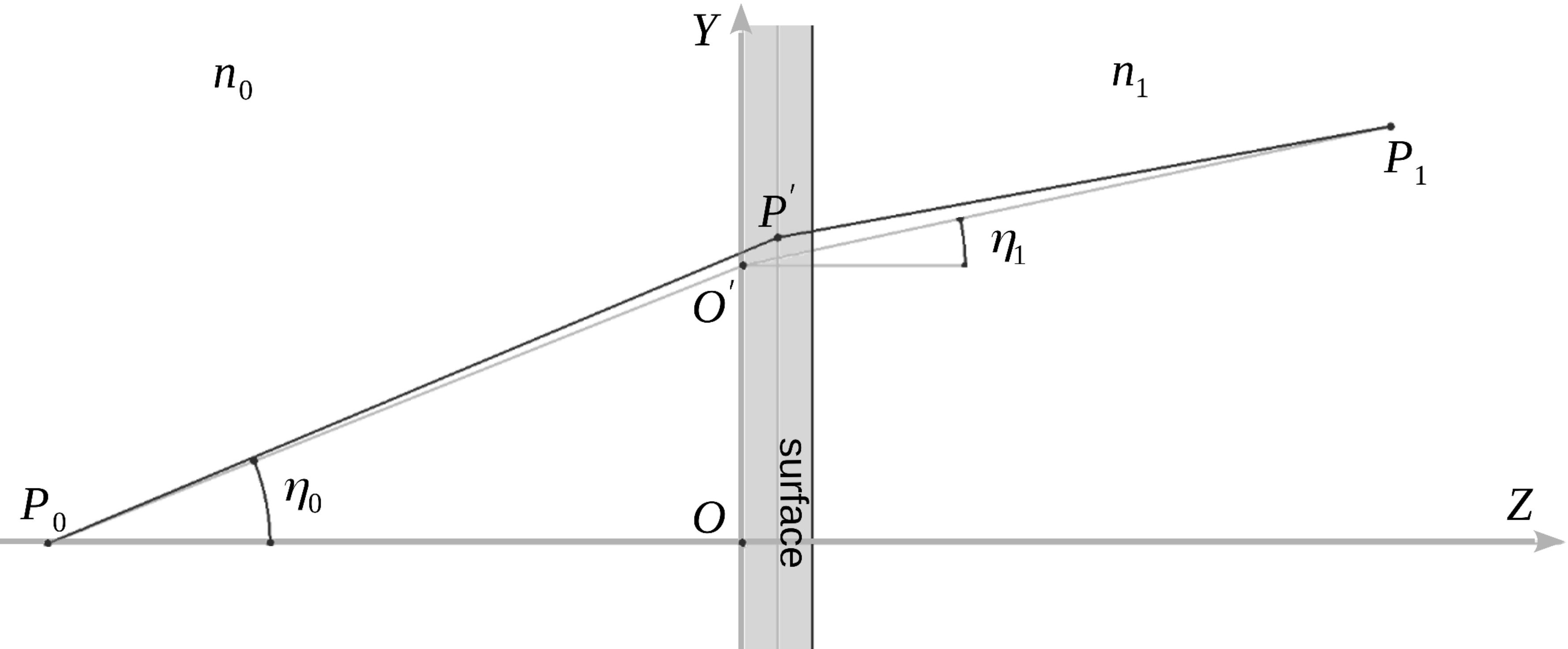}
	\caption{$YZ$ projection of the path of arbitrary ray through the refracting surface.}\label{fig:2}
\end{figure}
Figures~\ref{fig:1} and~\ref{fig:2} show, respectively, the $XZ$ and $YZ$ projections of the cylindrical surface of \eqref{eq:surface}, whose generatrix coincides with the $Y$ axis and the symmetry axis of the aperture stop. The optical path length between the object point $P_0$ and the image point $P_1$ for the ray crossing the surface in $P^\prime$ can be written as
\begin{eqnarray}\label{eq:OPL}
&&n_0 P_0 P^\prime + n_1 P^\prime P_1 =  
n_0 \sqrt{ (x^\prime-x_0)^2 + 
	y^{\prime2} + (z^\prime-z_0)^2 } + \nonumber\\
&&n_1 \sqrt{ (x^\prime-x_1)^2 + (y^\prime-y_1)^2 + (z^\prime-z_1)^2 }, 
\end{eqnarray}
where we have chosen $y_0=0$ without loss of generality due to the translation invariance. 

The chief ray goes along the vectors $\overrightarrow{P_0O^\prime}$ and $\overrightarrow{O^\prime P_1}$, where $O^\prime=(0,y^\prime,0)$. In the following we use the relations
\begin{eqnarray}
x_0 &=& - s_0 \sin\theta_0, \qquad x_1 = s_1 \sin\theta_1, \label{eq:relations0}\\
z_0 &=& - s_0 \cos\theta_0,  \qquad z_1 = s_1 \cos\theta_1, \label{eq:relations1}\\
s_i^2 &=& x_i^2+z_i^2, \quad i=0,1, \label{eq:si}
\end{eqnarray}
where $s_0$ and $s_1$ are the lengths of the $XZ$ projections of $\overrightarrow{P_0O^\prime}$ and $\overrightarrow{O^\prime P_1}$ vectors, respectively, and $\theta_0$ and $\theta_1$ are the angles for these vector projections with respect to the positive direction of $Z$ axis (positive angles are measured counterclockwise from $\overrightarrow{O^\prime Z}$). 

Since normal to the surface in \eqref{eq:surface} in the points 
along the $Y$ axis is parallel to the $Z$ axis, the Snell's law of refraction for the chief ray can be written as
\begin{eqnarray}
n_0 \sin\theta_0 = n_1 \sin\theta_1 \label{eq:SnellXZ}
\end{eqnarray}
in the $XZ$ plane and
\begin{eqnarray}
n_0 \sin\eta_0 = n_1 \sin\eta_1 \label{eq:SnellYZ}
\end{eqnarray}
in the $YZ$ plane, where $\eta_0$ and $\eta_1$ are the angles for the $YZ$ projections of the $\overrightarrow{P_0O^\prime}$ and $\overrightarrow{O^\prime P_1}$ vectors with respect to $\overrightarrow{O^\prime Z}$. 

The segment $P_0 P^\prime$ can be expressed as
\begin{eqnarray}\label{eq:segment0}
P_0 P^\prime &=& 
s_0 \left[ 1 + \frac{2x^\prime\sin\theta_0}{s_0} + \frac{x^{\prime2}}{s_0}\left( \frac{1}{s_0} + \frac{\cos\theta_0}{R} \right) \right.\nonumber\\
&+& \left. \frac{x^{\prime4}}{4s_0^2}\left( \frac{1}{R^2} + \alpha s_0\cos\theta_0 \right) + \frac{y^{\prime2}}{s_0^2} \right]^{1/2}.
\end{eqnarray}
Expanding the square root and retaining all terms through fourth order, we get
\begin{flalign}
&P_0 P^\prime = s_0 + x^\prime \sin\theta_0 + \frac{\Theta_0^x}{2}\left( \frac{x^{\prime2}}{s_0} - \frac{x^{\prime3}}{s_0^2}\sin\theta_0 \right) \nonumber\\
&+ \frac{1}{2}\left(\frac{y^{\prime2}}{s_0} - 
\frac{x^\prime y^{\prime2}}{s_0^2}\sin\theta_0\right)
- \frac{1}{8s_0^3} \left( x^{\prime4}\Theta_0^{xx} - 2x^{\prime2}y^{\prime2}\Theta_0^{xy} + y^{\prime4} \right),
\end{flalign}
where
\begin{flalign}\label{eq:Theta}
&\Theta_0^x = \cos\theta_0\left( \cos\theta_0 + \frac{s_0}{R} \right), 
\ \Theta_0^{xy} = 
2-3\cos^2\theta_0 - \frac{s_0}{R}\cos\theta_0,\nonumber\\
&\Theta_0^{xx} = 5\sin^4\theta_0 - 6\sin^2\theta_0\Theta_0 + \Theta_0^2 - \frac{s_0^2}{R^2} - \alpha s_0^3\cos\theta_0, \nonumber\\ 
&\Theta_0 = 1+\frac{s_0}{R}\cos\theta_0.
\end{flalign}

Using \eqref{eq:SnellYZ} and relation $\sin\eta\approx\tan\eta$ 
for small paraxial $\eta$, 
we express $y_1$ as
\begin{eqnarray}\label{eq:y1}
y_1 = y^\prime \left( 1 - \frac{n_0}{n_1}\frac{z_1}{z_0} \right),
\end{eqnarray}
where we ignored $O^\prime P^\prime$ with respect to $P_0P^\prime$. 
Using \eqref{eq:relations1}, we get
\begin{eqnarray}
y^\prime - y_1 = - \,\beta\, y^\prime
\end{eqnarray}
with
\begin{eqnarray}\label{eq:beta}
\beta = \frac{n_0 s_1\cos\theta_1}{n_1 s_0\cos\theta_0}.
\end{eqnarray}
Then $P^\prime P_1$ can be expressed as
\begin{eqnarray}
P^\prime P_1 &=& 
s_1 \left[ 1 - \frac{2x^\prime\sin\theta_1}{s_1} + \frac{x^{\prime2}}{s_1}\left( \frac{1}{s_1} - \frac{\cos\theta_1}{R} \right) \right.\nonumber\\
&+&\left. \frac{x^{\prime4}}{4s_1^2}\left( \frac{1}{R^2} - \alpha s_1\cos\theta_1 \right) + \frac{y^{\prime2}}{s_1^2}\beta^2 \right]^{1/2} \nonumber\\
&=&
s_1 - x^\prime \sin\theta_1 + \frac{\Theta_1^x}{2}\left(\frac{x^{\prime2}}{s_1} +
\frac{x^{\prime3}}{s_1^2}\sin\theta_1\right) \nonumber\\
&+& \frac{\beta^2}{2}\left(\frac{y^{\prime2}}{s_1} 
+ \frac{x^\prime y^{\prime2}}{s_1^2}\sin\theta_1 \right)
\nonumber\\
&-& \frac{1}{8s_1^3} \left( x^{\prime4}\Theta_1^{xx} - 2x^{\prime2}y^{\prime2}\beta^2\Theta_1^{xy} + y^{\prime4}\beta^4\right),
\end{eqnarray}
where
\begin{flalign}
&\Theta_1^x = \cos\theta_1\left( \cos\theta_1 - \frac{s_1}{R} \right), 
\ \Theta_1^{xy} = 2-3\cos^2\theta_1 + \frac{s_1}{R}\cos\theta_1,\nonumber\\
&\Theta_1^{xx} = 5\sin^4\theta_1 - 6\sin^2\theta_1\Theta_1 + \Theta_1^2 - \frac{s_1^2}{R^2} + \alpha s_1^3\cos\theta_1, \nonumber\\ 
&\Theta_1 = 1-\frac{s_1}{R}\cos\theta_1.
\end{flalign}

Now the optical path length in \eqref{eq:OPL} can be rewritten as
\begin{flalign}\label{eq:OPD}
&(n_0s_0 + n_1s_1) + x^\prime (n_0\sin\theta_0 - n_1\sin\theta_1) \nonumber\\
&+ \frac{x^{\prime2}}{2}\left( \frac{n_0}{s_0}\Theta_0^x + \frac{n_1}{s_1}\Theta_1^x \right)  -\frac{x^{\prime3}}{2}\left(n_0\frac{\sin\theta_0}{s_0^2}\Theta_0^x - n_1\frac{\sin\theta_1}{s_1^2}\Theta_1^x\right) \nonumber\\
&+ \frac{y^{\prime2}}{2}\left( \frac{n_0}{s_0} + \frac{n_1\beta^2}{s_1} \right) 
- \frac{x^\prime y^{\prime2}}{2}\left( \frac{n_0\sin\theta_0}{s_0^2} - \frac{n_1\beta^2\sin\theta_1}{s_1^2} \right) \nonumber\\
&- \frac{x^{\prime4}}{8} \left( \frac{n_0}{s_0^3}\Theta_0^{xx} + \frac{n_1}{s_1^3}\Theta_1^{xx} \right) 
+ \frac{x^{\prime2}y^{\prime2}}{4} \left( \frac{n_0}{s_0^3}\Theta_0^{xy} + \frac{n_1\beta^2}{s_1^3}\Theta_1^{xy} \right) \nonumber\\
&- \frac{y^{\prime4}}{8} \left(\frac{n_0}{s_0^3}+\frac{n_1\beta^4}{s_1^3}\right).
\end{flalign}

For the chief ray optical path length, using \eqref{eq:si}, at the same order of precision we have
\begin{eqnarray}\label{eq:chief}
&&n_0 \sqrt{x_0^2+z_0^2+y^{\prime2}} + n_1 \sqrt{x_1^2+z_1^2+(y^\prime-y_1)^2} \nonumber\\ 
&&= n_0s_0\sqrt{1+\frac{y^{\prime2}}{s_0^2}} + n_1s_1\sqrt{1+\frac{(y^\prime-y_1)^2}{s_1^2}} \nonumber\\ 
&&\approx (n_0s_0+n_1s_1) + \frac{n_0 y^{\prime2}}{2s_0} - \frac{n_0 y^{\prime4}}{8s_0^3} 
+ \frac{n_1 \beta^2y^{\prime2}}{2s_1} - \frac{n_1 \beta^4y^{\prime4}}{8s_1^3}.\nonumber\\
\end{eqnarray}
By subtracting \eqref{eq:chief} from \eqref{eq:OPD} we get the wave aberration 

\begin{flalign}
&\Phi^{(4)} = 
 \alpha_d \,x^\prime + \alpha_{a1}\frac{x^{\prime2}}{2} + \alpha_{a2}\frac{y^{\prime2}}{2} + \alpha_{c1}\frac{x^{\prime3}}{3} + \alpha_{c2}\,x^\prime y^{\prime2} \nonumber\\
&+~ \alpha_{e1}\frac{x^{\prime4}}{4} +  \alpha_{e2}\frac{y^{\prime4}}{4} +
\alpha_{e3}\frac{x^{\prime2}y^{\prime2}}{2},
\label{eq:wave_aber}
\end{flalign}
where the terms with the coefficients $\alpha_{a2}$ and $\alpha_{e2}$ vanish due to the approximations used, while the rest coefficients read
\begin{eqnarray}\label{eq:abers_first}
\alpha_{e1} &=& - \frac{1}{2} \left( \frac{n_0}{s_0^3}\Theta_0^{xx} + \frac{n_1}{s_1^3}\Theta_1^{xx} \right), \\  
\alpha_{e3} &=& \frac{1}{2} \left( \frac{n_0}{s_0^3}\Theta_0^{xy} + \frac{n_1\beta^2}{s_1^3}\Theta_1^{xy} \right), \\
\alpha_{c1} &=& -\frac{3}{2}\left(n_0\frac{\sin\theta_0}{s_0^2}\Theta_0^x - n_1\frac{\sin\theta_1}{s_1^2}\Theta_1^x\right), \\
\alpha_{c2} &=&  -\frac{1}{2}\left( \frac{n_0\sin\theta_0}{s_0^2} - \frac{n_1\beta^2\sin\theta_1}{s_1^2} \right), \\
\alpha_{a1} &=& \frac{n_0}{s_0}\Theta_0^x + \frac{n_1}{s_1}\Theta_1^x, 
\\ 
\alpha_d &=& n_0\sin\theta_0 - n_1\sin\theta_1 \label{eq:abers_last}
\end{eqnarray}
with the sub-indices $e$, $c$, $a$ and $d$ referring to the elliptic aberration, coma, astigmatism and distortion, respectively. 

Using the approximate formulas
\begin{eqnarray}
\delta_{1x} = \frac{s_1}{n_1}\frac{\partial\Phi^{(4)}}{\partial x^\prime}, \qquad 
\delta_{1y} = \frac{s_1}{n_1}\frac{\partial\Phi^{(4)}}{\partial y^\prime},
\end{eqnarray}
for the components of transverse aberration we have
\begin{flalign}
&\delta_{1x} = \frac{s_1}{n_1}\left[(\alpha_{e1} x^{\prime2} + \alpha_{e3} y^{\prime2})x^\prime + (\alpha_{c1} x^{\prime2} + \alpha_{c2} y^{\prime2}) + \alpha_{a1} x^\prime + \alpha_d\right], \nonumber\\
&\delta_{1y} = \frac{s_1}{n_1}\left[(\alpha_{e3} x^{\prime2} + \alpha_{e2} y^{\prime2})y^\prime + 2\alpha_{c2} x^\prime y^{\prime} + \alpha_{a2} y^\prime \right].
\end{flalign}

We remark that the \verb+Wolfram Mathematica+ software~\cite{Mathematica} was used for partial checks. 

\subsection{Evaluation of aberration coefficients}

The aberration functions in Eqs.~(\ref{eq:abers_first})--(\ref{eq:abers_last}) can be evaluated as follows. For the terms representing coma we replace each cosine by one and each sine by its angle. Using~\eqref{eq:SnellXZ}, we have
\begin{eqnarray}
\alpha_{c2} = - \frac{n_0\theta_0}{2s_0^2} \left( 1 - \frac{n_0^2}{n_1^2} \right).
\end{eqnarray}
Then, using the following relation 
for a surface of cylinder with the radius $R$
\begin{eqnarray}
\frac{n_1-n_0}{R} = \frac{n_1}{z_1} - \frac{n_0}{z_0}
\end{eqnarray}
and \eqref{eq:relations1}, we get the formula
\begin{eqnarray}\label{eq:surface_power}
n_0 \left( \frac{1}{s_0} + \frac{1}{R} \right) = - n_1 \left( \frac{1}{s_1} - \frac{1}{R} \right)
\end{eqnarray}
that allows us to rewrite $\alpha_{c1}$ as
\begin{eqnarray}
\alpha_{c1} = - \frac{3}{2}\frac{n_0\theta_0}{s_0^2} \left( 1+\frac{s_0}{R} \right) \left( 1 + \frac{n_0}{n_1}\frac{s_0}{s_1} \right),
\end{eqnarray}
which in case of $R\to\infty$ reduces to $3\alpha_{c2}$.

Again approximating cosines by unity and sines by their angles, using \eqref{eq:surface_power} and the leading order approximation of the Snell's law
\begin{eqnarray}\label{eq:Snell_approx}
n_0\theta_0=n_1\theta_1,
\end{eqnarray}
for the elliptic aberration coefficients we have
\begin{flalign}
&\alpha_{e1} = -\frac{1}{2}\left[ \frac{\kappa}{R^3}(n_1-n_0) + b \right] - \frac{n_0}{2s_0^3} \left( 1+\frac{s_0}{R} \right)^2 \left( 1+\frac{n_0s_0}{n_1s_1} \right), \label{eq:alfa_e1}\\
&\alpha_{e3} = -\frac{n_0}{2s_0^3} \left( 1+\frac{s_0}{R} \right) \left( 1-\frac{n_0^2}{n_1^2} \right).
\end{flalign}

Using the expansion of $\beta$ to the second order in $\theta_i$ angles, 
\begin{eqnarray}
\beta = \frac{n_0 s_1}{n_1 s_0} \left[ 1 + \frac{\theta_0^2}{2}\left( 1-\frac{n_0^2}{n_1^2} \right) \right],
\end{eqnarray}
the coefficients $\alpha_{ai}$ can be expressed 
as
\begin{eqnarray}
\alpha_{a1} &=& -\frac{n_0\theta_0^2}{s_0} \left( 1 + \frac{n_0s_0}{n_1s_1} \right) - \frac{n_0\theta_0^2}{2R} \left( 1 - \frac{n_0}{n_1} \right). 
\end{eqnarray}

In consideration of $\alpha_d$ the first order approximation of the Snell's law in \eqref{eq:Snell_approx} is not sufficient. In order to preserve the precise direction from which the chief ray approaches the stop (and the object and image planes), it is necessary to expand sines and tangents to higher orders~\cite{Schroeder}. From Eqs.~(\ref{eq:relations0})--(\ref{eq:relations1}) we have $\tan\theta_0=x_0/z_0$ and $\tan\theta_1=x_1/z_1$. To third order
\begin{eqnarray}
\theta_0 = \frac{x_0}{z_0} - \frac{1}{3}\theta_0^3, \qquad \theta_1 = \frac{x_1}{z_1} - \frac{1}{3}\theta_1^3.
\end{eqnarray}
We now substitute this to $\alpha_d$ expanded to third order and find
\begin{eqnarray}
\alpha_d = \left(\frac{n_0x_0}{z_0} - \frac{n_1x_1}{z_1}\right) - n_0\frac{\theta_0^3}{2}\left(1-\frac{n_0^2}{n_1^2}\right),
\end{eqnarray}
where the terms in the first brackets cancel each other.

The above aberration coefficients for the considered refracting surface are summarized in the Table~\ref{tab:aberrations1} (second column), where $\beta=(n_0s_1)/(n_1s_0)$ and the following notations are used
\begin{eqnarray}
\Gamma &=& \frac{1}{s_0} \left( 1 + \frac{n_0}{n_1}\frac{s_0}{s_1} \right), \\
\Gamma_R &=& \frac{1}{s_0} \left( 1 + \frac{s_0}{R} \right).
\end{eqnarray}
\begin{table}
	\caption{Aberration coefficients for a cylindrical surface}\label{tab:aberrations1}
	\begin{tabular}{l||l|l}
		\hline
		$\!\alpha$ & Refracting surface of \eqref{eq:surface} & Reflecting surface,\,(\ref{eq:surface2}) \\
		\hline
		\hline
		$\!\alpha_{e1}$ & $\!-\frac{1}{2}\!\left[ \frac{\kappa}{R^3}(n_1-n_0) + b + n_0\Gamma\Gamma_R^2 \right]\!$ 
		& $-\frac{n_0}{R^3}\left[ \kappa\! +\! \left( \frac{\mu+1}{\mu-1} \right)^2 \right]\! - \frac{b}{2}\!$ \\
		$\!\alpha_{e3}$ & $-\frac{n_0}{2s_0^2} \left( 1-\frac{n_0^2}{n_1^2}  \right) \Gamma_R$ 
		& $-\frac{n_0}{s_0^2R}\frac{\mu+1}{\mu-1}$
		\\
		\hline
		$\!\alpha_{c1}$ & $-\frac{3}{2}n_0\theta_0 \Gamma\Gamma_R$  
		& $-3\frac{n_0\theta_0}{R^2}\frac{\mu+1}{\mu-1}$ \\
		$\!\alpha_{c2}$ & $- \frac{n_0\theta_0}{2s_0^2} \left( 1 - \frac{n_0^2}{n_1^2} \right)$ & 0 \\
		\hline
		$\!\alpha_{a1}$ & $-n_0\theta_0^2 \left( \Gamma + \frac{n_1-n_0}{2n_1R} \right)$ 
		& $-\frac{n_0\theta_0^2}{R}$\\ 
		\hline
		$\!\alpha_d$ & $-\frac{n_0\theta_0^3}{2}\left(1-\frac{n_0^2}{n_1^2}\right)$ & 0 \\
		\hline
	\end{tabular}
\end{table}
We notice that the expressions for $A_0$, $A_2$, $A_3$ and $A_1$ from Table~5.1 of Ref.~\cite{Schroeder} correspond to $\alpha_d$, $\alpha_{c1}/3$, $\alpha_{e1}/4$ and the first term in the expression for $\alpha_{a1}/2$, respectively, taking into account the correspondence of $s_0\to-s$ and $s_1\to s^\prime$. 

\subsection{Aberrations for a mirror surface}

Consider now a reflecting cylindrical surface that is described by the equation
\begin{eqnarray}\label{eq:surface2}
z = -\frac{x^2}{2R} - \frac{x^4}{8} \left( \frac{1+\kappa}{R^3} + \frac{b}{2n_0} \right) \equiv 
-\frac{x^2}{2R} - \frac{\alpha x^4}{8},
\end{eqnarray}
where 
$n_0$ is the index of refraction before the surface. For the law of reflection in the $XZ$ plane we have $\theta_1 = \pi-\theta_0$. Then up to the third order in $\theta_0$
\begin{eqnarray}
\cos\theta_0 =& -\cos\theta_1 &= 1-\frac{\theta_0^2}{2},\label{eq:X1} \\
\sin\theta_0 =& \sin\theta_1 &= \theta_0-\frac{\theta_0^3}{6}
\end{eqnarray} 
The results for the aberrations in Eqs.~(\ref{eq:abers_first})--(\ref{eq:abers_last}) remain correct after the substitution of 
\begin{eqnarray}
R\to-R, \qquad \alpha\to-\alpha. \label{eq:X3}
\end{eqnarray}

Using Eqs.~(\ref{eq:X1})--(\ref{eq:X3}), we have
\begin{flalign}
&\Theta_i^x = \left(1 - \frac{s_i}{R}\right) - \frac{\theta_0^2}{2}\left( 2-\frac{s_i}{R} \right), 
~\Theta_i^{xy} = -\left(1 - \frac{s_i}{R}\right) + \frac{\theta_0^2}{2}\left( 6-\frac{s_i}{R} \right)\!,\\
&\Theta_i^{xx} = \left( 1 - \frac{s_i}{R} \right)^2 - \frac{s_i^2}{R^2} + \alpha s_i^3 - \frac{\theta_0^2}{2} \left[ 2\left( 1 - \frac{s_i}{R} \right) \left( 6 - \frac{s_i}{R} \right) + \alpha s_i^3 \right]
\end{flalign}
for $i=0,1$. 

For the mirror surface the function $\beta$ in \eqref{eq:beta} reduces to the transverse (lateral) magnification
\begin{eqnarray}
\mu = -\frac{s_1}{s_0},
\end{eqnarray}
which is negative for the considered concave mirror. However, for a convex mirror with either virtual object or virtual image the respective distance should be taken with the minus sign that changes the sign of $\mu$.

From the law of reflection in the $YZ$ plane
\begin{eqnarray}\label{eq:eta_reflection_law}
\eta_1 = \pi - \eta_0,
\end{eqnarray}
using Eqs.~(\ref{eq:relations1}) and (\ref{eq:X1}),
the analog of \eqref{eq:y1} writes as
\begin{eqnarray}
y_1 = y^\prime\left(1+\frac{z_1}{z_0}\right) \approx y^\prime(1-\mu).
\end{eqnarray}

Using the law of reflection
\begin{eqnarray}
\frac{1}{s_1} + \frac{1}{s_0} = \frac{2}{R},
\end{eqnarray}
we have 
\begin{eqnarray}
\frac{1}{s_0}\left( 1-\frac{s_0}{R} \right) = - \frac{1}{s_1}\left( 1-\frac{s_1}{R} \right) = \frac{1}{R} \, \frac{\mu+1}{\mu-1}
\end{eqnarray}
and derive the expressions for $\alpha_i$ shown in the third column of Table~\ref{tab:aberrations1}. 
The distortion coefficient $\alpha_d$ is vanishing same as for a mirror of revolution~\cite{Schroeder}. However, in the case of cylindrical mirror the second coma is also vanishing: $\alpha_{c2}=0$. 
Hence, in this case only two additional aberrations left with respect to the case of axially symmetric surface, and these two new degrees of freedom modify the spherical aberration. 
Possibilities for compensation of the modified spherical aberration in the systems of cylindrical mirrors are discussed in section~\ref{sec:multisurface_systems}.

\subsection{Aberrations for displaced stop}

\begin{figure}[ht!]
	\centering\includegraphics[width=7.2cm]{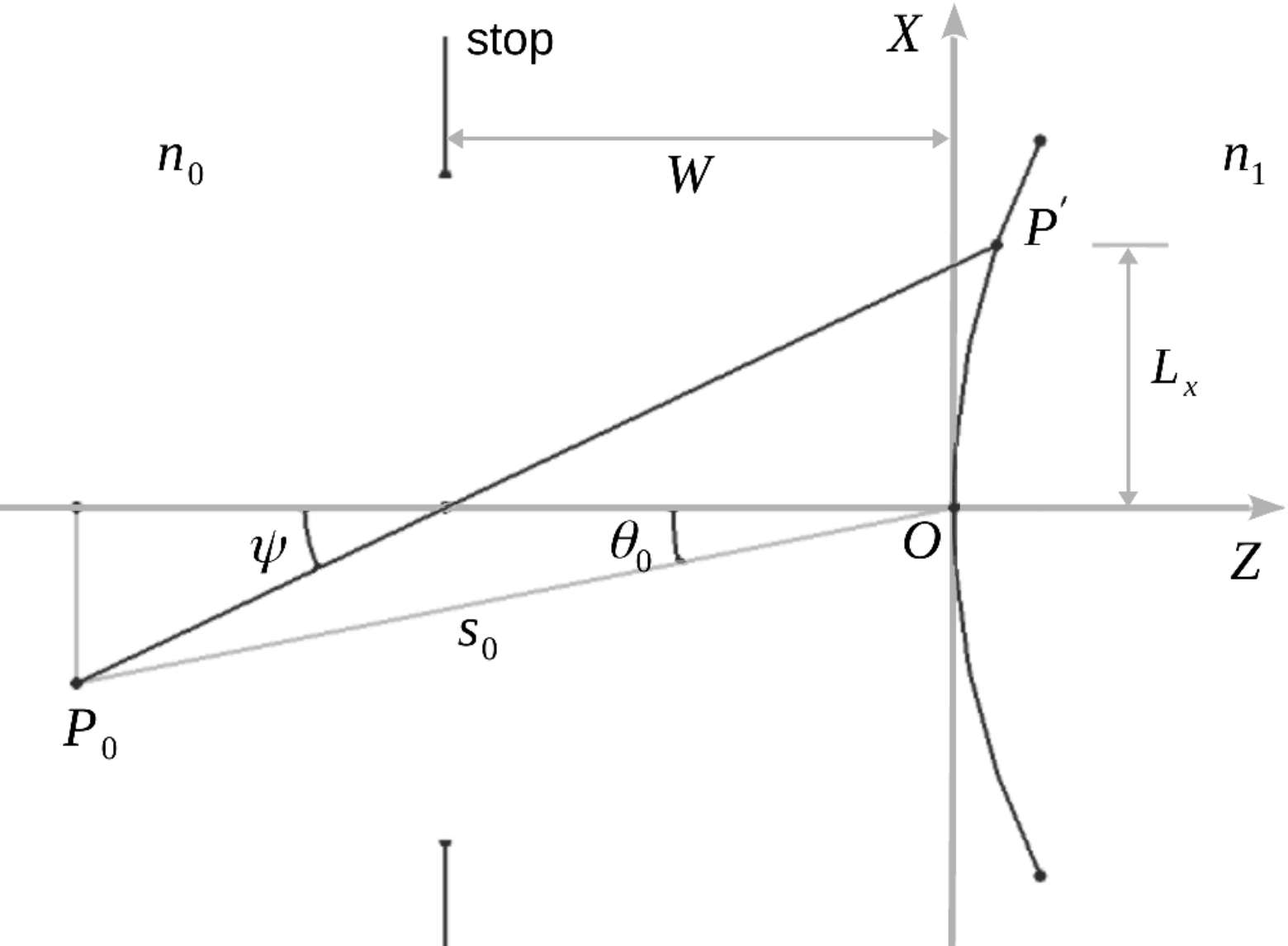}\vspace{6mm}
	\centering\includegraphics[width=6cm]{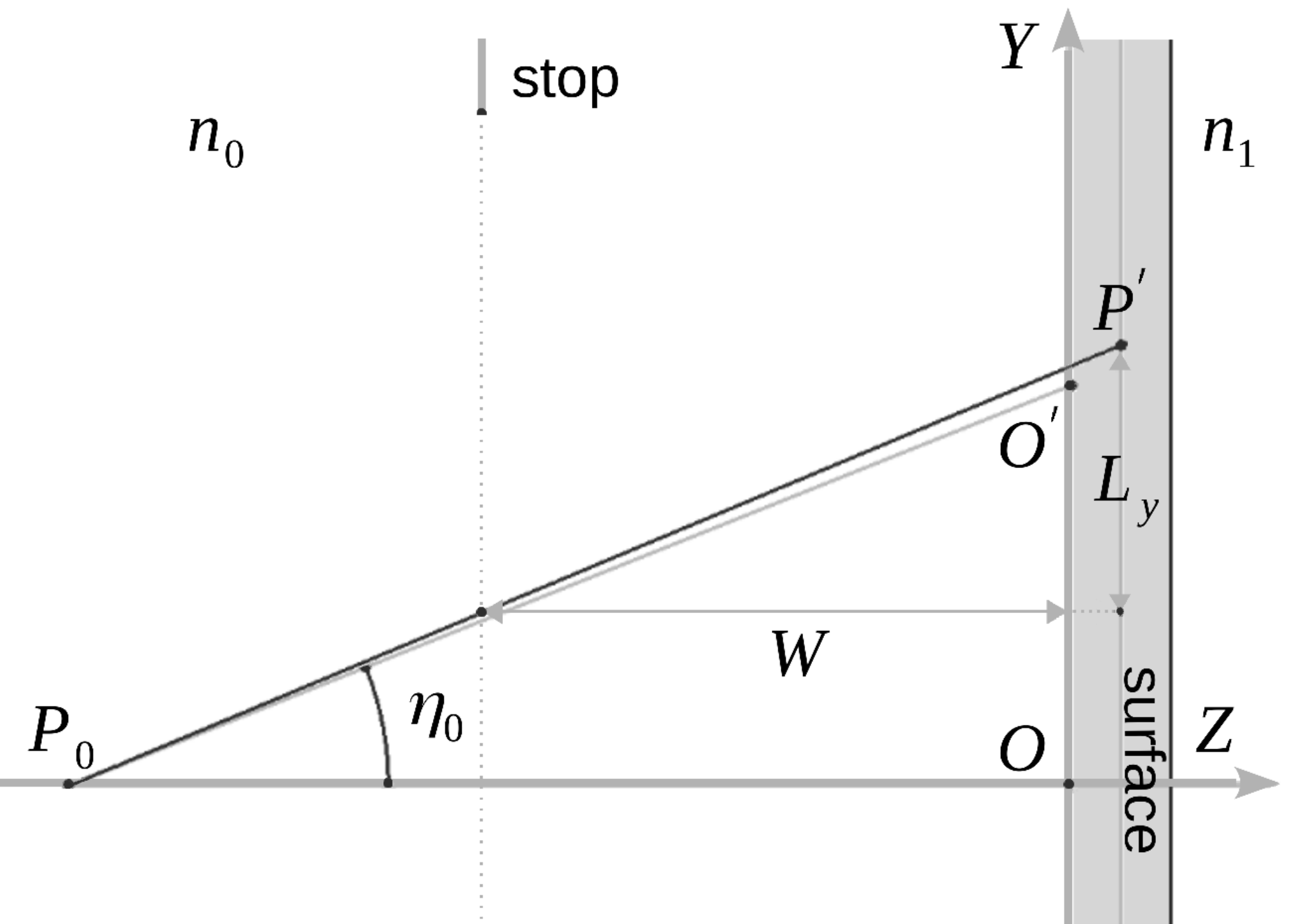}
	\caption{{\it Upper} [{\it Lower}]: portion of Fig.~\ref{fig:1} [\ref{fig:2}] with aperture stop displaced from surface by a distance $W$ along the $Z$ axis. The entrance pupil is at the stop. The chief ray 
	makes angle $\psi$ [$\chi\approx\eta_0$] with the $Z$ axis and intersects the surface at height $L_x$ [$L_y$].}\label{fig:stop}
\end{figure}

Consider a single optical surface and the aperture stop displaced from it by a distance $W$, 
as shown in Fig.~\ref{fig:stop}. 
The chief ray, which is now $P_0P^\prime$, is directed towards the pupil's center and intersects the surface at heights $L_x$ and $L_y$ in the $XZ$ and $YZ$ planes, respectively, and the wave aberration writes as 
$\Phi(x^\prime+L_x,y^\prime+L_y)$. 
However, taking into account the fact that the ray behavior in the $XZ$ plane is the same in the discussed cylindrical OS as in the corresponding rotationally symmetric OS with the same $XZ$ profile, we consider the $\Phi(x^\prime+L_x,y^\prime)$ part first to simplify further comparisons to the classical results~\cite{Schroeder}.

\subsubsection{$L_y=0$ part of the aberration function}

Making the replacement of $x^\prime\to x^\prime+L_x$ in \eqref{eq:wave_aber} and collecting terms in various powers of $x^\prime$ and $y^\prime$, we get  
\begin{flalign}\label{eq:Phi_stop}
\Phi^{(4)}(x^\prime+L_x,y^\prime) &= A_{01}x^\prime + A_{11}x^{\prime2} + A_{12} y^{\prime2} + A_{21}x^{\prime3} + A_{22} x^\prime y^{\prime2} \nonumber\\ 
&+ A_{31}x^{\prime4} + A_{32}y^{\prime4} 
+ A_{33}x^{\prime2}y^{\prime2} + C_0,
\end{flalign}
where $C_0$ is the constant that is irrelevant in the following and the coefficients $A_{ij}$ write as
\begin{flalign}
A_{01} &= \alpha_{d} + \alpha_{a1}L_x + \alpha_{c1}L_x^2 + \alpha_{e1}L_x^3, \label{eq:B0aber}\\
A_{11} &= \frac{\alpha_{a1}}{2} + \alpha_{c1}L_x + \frac{3\alpha_{e1}}{2}L_x^2, \quad 
A_{12} = 
\alpha_{c2}L_x + \frac{\alpha_{e3}}{2}L_x^2, \\
A_{21} &= \frac{\alpha_{c1}}{3} + \alpha_{e1}L_x, \quad
A_{22} = \alpha_{c2} + \alpha_{e3}L_x, \\
A_{31} &= \frac{\alpha_{e1}}{4}, \quad  A_{32} = 0,
\quad  A_{33} = \frac{\alpha_{e3}}{2}. \label{eq:B3aber}
\end{flalign}
From Eqs.~(\ref{eq:B0aber})--(\ref{eq:B3aber}) we have the following stop-shift relations (they can be applied also to a system made up of many surfaces):
\begin{itemize}
	\item elliptic aberration coefficients $A_{3i}$ ($i=1,2,3$) are independent of the stop position;
	\item if $\alpha_{ej} = 0$ then the aberration coefficient $A_{2j}$ is independent of the stop position;
	\item if $\alpha_{ej} = \alpha_{cj} = 0$ then the coefficient $A_{1j}$ is also independent of the stop position.
\end{itemize}

From Fig.~\ref{fig:stop} one can get the relations
\begin{eqnarray}
L_x &=& W\psi + \mathcal{O}\left(\psi^3\right), \label{eq:relat1}\\
\theta_0 &=& W\psi\left( \frac{1}{W} - \frac{1}{s_0} \right) + \mathcal{O}\left(\psi^3\right), \label{eq:relat2}
\end{eqnarray} 
where $W>0$ is the distance between the surface and the stop (entrance pupil is in front of the surface and coincides with the stop), and $\psi$ is the angle between the chief ray and the $Z$ axis.
However, if the entrance pupil and stop are behind the surface, 
the relations analogous to Eqs.~(\ref{eq:relat1}) and (\ref{eq:relat2}) read as
\begin{eqnarray}
L_x &=& -W\psi + \mathcal{O}\left(\psi^3\right), \\
\theta_0 &=& W\psi\left( \frac{1}{W} + \frac{1}{s_0} \right) + \mathcal{O}\left(\psi^3\right), \label{eq:relat4}
\end{eqnarray}
where $W>0$ is the distance between the surface and the entrance pupil.

Using the above definitions we find the following algebraic relations
\begin{flalign}\label{eq:formula1}
\theta_0 + L_x\Gamma_R &= W\psi \left( \frac{1}{W}\pm\frac{1}{R} \right),\\
2\theta_0^2 + 3\theta_0L_x\Gamma_R + L_x^2\Gamma_R^2 &= (W\psi)^2 \left( \frac{1}{W}\pm\frac{1}{R} \right) \left( \frac{2}{W}\pm\frac{1}{R}\mp\frac{1}{s_0} \right),\\
2\theta_0^2 + 6\theta_0L_x\Gamma_R + 3L_x^2\Gamma_R^2 &= (W\psi)^2 \nonumber\\
 &\times \left[ 3\left( \frac{1}{W}\pm\frac{1}{R} \right)^2 - \left( \frac{1}{W}\mp \frac{1}{s_0} \right)^2 \right]. \label{eq:formula3}
\end{flalign}

We remark that Eqs.~(\ref{eq:relat1})--(\ref{eq:formula3}) are valid for small $\theta_0$ and $\psi$. In general, for paraxial rays in a multisurface OS the values of these angles for any particular surface are either both close to 0 or both close to $\pi$. However, in the later case we can redefine for these evaluations the angles as $\theta_0\to\pi-\theta_0$ and $\psi\to\pi-\psi$ to bring their values close to zero and keep Eqs.~(\ref{eq:relat1})--(\ref{eq:formula3}) correct. Alternatively, one can choose the direction of local $Z_i$ axis towards the $i$th surface from the side of incoming ray.

From the results for $\alpha_i$ in Table~\ref{tab:aberrations1} and Eqs.~(\ref{eq:formula1})--(\ref{eq:formula3}), we derive the expressions for the coefficients $A_{ij}$ shown in Tables~\ref{tab:aberrations21} and \ref{tab:aberrations22}, where the upper and lower signs correspond to the stop located in front of (see Fig.~\ref{fig:stop}) and behind the surface, respectively.

\begin{table}
	\caption{Aberration coefficients for the refracting cylindrical  surface of \eqref{eq:surface} 
		with displaced stop and $L_y=0$}\label{tab:aberrations21}
	\begin{tabular}{l|l}
		\hline
		$A_{31}$~ & ~$-\frac{n_0}{8}\left( \frac{\kappa}{R^3}\frac{n_1-n_0}{n_0} + \frac{b}{n_0} + \Gamma\Gamma_R^2 \right)$ \\
		\hline
		$A_{33}$ & ~$-\frac{n_0}{4s_0^2} \left( 1-\frac{n_0^2}{n_1^2} \right) \Gamma_R$ \\
		\hline
		$A_{21}$ & ~$- \frac{n_0}{2}(W\psi) \left[ \pm\frac{\kappa}{R^3}\frac{n_1-n_0}{n_0} \pm \frac{b}{n_0} + \left( \frac{1}{W}\pm \frac{1}{R} \right)\Gamma\Gamma_R \right]$ \\
		\hline
		$A_{22}$ & ~$- \frac{n_0}{2s_0^2} (W\psi) \left( \frac{1}{W}\pm \frac{1}{R} \right) \left( 1 - \frac{n_0^2}{n_1^2} \right)$ \\
		\hline
		$A_{11}$ & ~$-\frac{n_0}{4}(W\psi)^2 \left[ 3\left( \frac{1}{W}\pm \frac{1}{R} \right)^2 - \left( \frac{1}{W}\mp \frac{1}{s_0} \right)^2 \right]\Gamma$ \\
		& ~$-\frac{n_0}{4}(W\psi)^2 \left[ \left( \frac{1}{W}\mp \frac{1}{s_0} \right)^2\frac{n_1-n_0}{n_1R} + \frac{3\kappa}{R^3}\frac{n_1-n_0}{n_0} +\frac{3b}{n_0}\right]$ \\
		\hline
		$A_{12}$ 
		& ~$-\frac{n_0}{2s_0^2} (W\psi)^2 \left( \frac{1}{2R} \pm\frac{1}{W}-\frac{1}{2s_0} \right) \left( 1 - \frac{n_0^2}{n_1^2} \right)$ \\
		\hline
		$A_{01}$ & ~$-\frac{n_0}{2} (W\psi)^3 \left\{ \left( \frac{1}{W} \pm \frac{1}{R} \right) \left( \frac{1}{R} \pm\frac{2}{W} -\frac{1}{s_0} \right) \Gamma \right.$ \\
		& ~$+\left( \frac{1}{W} \mp \frac{1}{s_0} \right)^2 \left[ \left( \frac{1}{W} \mp \frac{1}{s_0} \right) \left( 1 - \frac{n_0^2}{n_1^2} \right) \pm \frac{n_1-n_0}{n_1R} \right]$ \\
		& ~$\left. \pm \left( \frac{\kappa}{R^3}\frac{n_1-n_0}{n_0} +\frac{b}{n_0} \right) \right\}$ \\
		\hline
	\end{tabular}
\end{table}

\begin{table}
	\caption{Aberration coefficients for the reflecting cylindrical surface of \eqref{eq:surface2} 
		with displaced stop and $L_y=0$}\label{tab:aberrations22}
	\begin{tabular}{l|l}
		\hline
		$A_{31}$~ & ~$-\frac{n_0}{4R^3} \left[ \kappa + \left( \frac{\mu+1}{\mu-1} \right)^2 \right] - \frac{b}{8}$ \\
		\hline
		$A_{33}$ & ~$-\frac{n_0}{2s_0^2R}\, \frac{\mu+1}{\mu-1}$ \\ 
		\hline
		$A_{21}$ & 
		~$-\frac{n_0}{R^2} (W\psi) \left[ \left( \frac{1}{W}\mp \frac{1}{R} \right) \frac{\mu+1}{\mu-1} \pm \frac{\kappa}{R} \right] \mp \frac{b}{2} (W\psi)$\\
		\hline
		$A_{22}$ & ~$\mp\frac{n_0}{2s_0^2R}\left(W\psi\right) \frac{\mu+1}{\mu-1}$ \\
		\hline
		$A_{11}$ & 
		~$-\frac{n_0}{2R} (W\psi)^2 \left[ 3\left( \frac{1}{W}\mp \frac{1}{R} \right)^2 - 2\left( \frac{1}{W}\mp \frac{1}{s_0} \right)^2 + \frac{3\kappa}{R^2} \right]$ \\
		& ~$- \frac{3b}{4} (W\psi)^2$ \\
		\hline
		$A_{12}$ & ~$
		-\frac{n_0}{2s_0^2R}\left(W\psi\right)^2 \frac{\mu+1}{\mu-1}$ \\
		\hline
		$A_{01}$ & 
		~$\mp\frac{n_0}{2R} (W\psi)^3 \left[ 3\left( \frac{1}{W}\mp \frac{1}{R} \right)^2 - \left( \frac{1}{W} \mp \frac{1}{s_0} \right)^2 \right]$ \\
		& 
		~$\mp\frac{n_0}{2R^3} (W\psi)^3 \left[ \kappa - \left( \frac{\mu+1}{\mu-1} \right)^2 \right] \mp \frac{b}{2} (W\psi)^3$ \\
		\hline
	\end{tabular}
\end{table}

The coefficients $A_{21}$ and $A_{31}$ are equal to $B_2$ and $B_3$ from Table~5.5 of Ref.~\cite{Schroeder}, respectively, taking into account that $W$ and $s_0$ are positive in our definitions.

By taking $A_{11}=0$ or $A_{12}=0$ one can locate the tangential or sagittal astigmatic images as discussed for the axially symmetric case in Ref.~\cite{Schroeder}.

\subsubsection{Rest of the terms}

From Eqs.~(\ref{eq:wave_aber}) and (\ref{eq:Phi_stop}) we have 
\begin{flalign}\label{eq:Phi_residue}
&\Phi^{(4)}(x^\prime+L_x,y^\prime+L_y) - \Phi^{(4)}(x^\prime+L_x,y^\prime) =  B_{01}x^\prime + B_{02}y^\prime \nonumber\\ 
&+ B_{11}x^{\prime2} + B_{12} y^{\prime2} + B_{13} x^\prime y^\prime + B_{22}y^{\prime3} + B_{23} x^{\prime2}y^\prime + C,
\end{flalign}
where $C$ is the new irrelevant constant and the coefficients $B_{ij}$ write as
\begin{flalign}
B_{01} &= \left( \alpha_{c2} + \alpha_{e3}L_x \right) L_y^2,  \quad B_{02} = \left( 
2\alpha_{c2} + \alpha_{e3}L_x 
\right)L_xL_y, \\
B_{11} &= \frac{\alpha_{e3}}{2}L_y^2,  \quad  B_{12} = 0, 
\quad  B_{13} = 2\left( \alpha_{c2} + \alpha_{e3}L_x \right) L_y, \\
B_{22} &= 0,
\quad  B_{23} = \alpha_{e3}L_y.
\end{flalign}
All of the $B_{ij}$ are stop-shift dependent.

In the $YZ$ plane
, denoting the angle $OP_0P^\prime$ in Fig.~\ref{fig:2} by $\chi$, we obtain
\begin{eqnarray}
	L_y = W\tan\chi = W\chi \left( 1+\frac{\chi^2}{3} \right) + \mathcal{O}(\chi^5),
	\label{eq:relat_chi}
\end{eqnarray}
where $\chi\approx\eta_0$ 
and the comment below \eqref{eq:formula3} is also applicable. 
The final expressions for $B_{ij}$ are shown in Tables~\ref{tab:aberrations_B_1} and \ref{tab:aberrations_B_2}. The upper and lower signs in these tables and in \eqref{eq:relat_chi} correspond to the entrance pupil located in front of the surface and behind it, respectively.
\begin{table}
	\caption{Aberration coefficients $B$ for the refracting cylindrical surface of \eqref{eq:surface} in case of displaced stop}\label{tab:aberrations_B_1}
	\begin{tabular}{l|l}
		\hline
		$B_{23}$ & ~$-\frac{n_0(W\chi)}{2s_0^2} \left( 1 - \frac{n_0^2}{n_1^2} \right)  \Gamma_R$ 
		\\
		\hline
		$B_{11}$ & ~$-\frac{n_0(W\chi)^2}{4s_0^2} \left( 1 - \frac{n_0^2}{n_1^2} \right) \Gamma_R$ 
		\\
		\hline
		$B_{13}$ & ~$-\frac{n_0W^2}{s_0^2}\psi\chi \left( 1 - \frac{n_0^2}{n_1^2} \right) \left( \frac{1}{W}\pm\frac{1}{R} \right)$ 
		\\
		\hline
		$B_{01}$ & ~$-\frac{n_0}{2s_0^2} W^3\psi\chi^2 \left( 1 - \frac{n_0^2}{n_1^2} \right) \left( \frac{1}{W}\pm\frac{1}{R} \right)$ 
		\\
		\hline
		$B_{02}$ 
		& ~$- \frac{n_0W^3}{2s_0^2} \psi^2\chi \left( 1 - \frac{n_0^2}{n_1^2} \right) \left[ \pm2
		\left( \frac{1}{W}\mp\frac{1}{s_0} \right) + \Gamma_R \right]$ \\
		\hline
	\end{tabular}
\end{table}

\section{Multisurface systems}\label{sec:multisurface_systems}

\subsection{Generic discussion}

Consider OS containing more then one optical surface. Let index $i$ numerate the surfaces along any incoming ray. For the $i$th surface the object $P_i$ and image $P_i^\prime$ points are located at distances $s_{0i}$ and $s_{1i}$, respectively, from the surface. The wave aberration for the system~is
\begin{eqnarray}
\Phi_s = \sum_{i=1}^f \Phi_i,
\end{eqnarray}
where $f$ refers to the final surface. At the fourth order, each $\Phi_i^{(4)}$ can be replaced by the sum of the expressions given in Eqs.~(\ref{eq:Phi_stop}) and (\ref{eq:Phi_residue}), with the appropriate $x^\prime$ and $y^\prime$ at each surface denoted by $x_i$ and $y_i$, respectively. Correspondingly, we denote the refractive indices $n_{0i}$ and $n_{1i}$, the distance $W_i$ from the $i$th surface to its entrance pupil, the angles $\psi_i$, $\chi_i$, etc.

\begin{table}
	\caption{Aberration coefficients $B$ for the reflecting cylindrical surface of \eqref{eq:surface2} in case of displaced stop}\label{tab:aberrations_B_2}
	\begin{tabular}{l|l}
		\hline
		$B_{23}$ 
		& ~$-\frac{n_0(W\chi)}{
			s_0^2R}\,\frac{\mu+1}{\mu-1}$ \\
		\hline
		$B_{11}$ 
		&	~$-\frac{n_0(W\chi)^2}{
			2s_0^2R}\,\frac{\mu+1}{\mu-1}$ \\
		\hline
		$B_{13}$ 
		& ~$\mp\frac{2n_0}{s_0^2R}W^2\psi\chi\, \frac{\mu+1}{\mu-1}$ \\
		\hline
		$B_{01}$ 
		& 
		~$\mp\frac{n_0}{s_0^2R}W^3\psi\chi^2 \frac{\mu+1}{\mu-1}$ \\
		\hline
		$B_{02}$ 
		& ~$
		- n_0\frac{W^3}{s_0^2R}\psi^2\chi \frac{\mu+1}{\mu-1}$ \\
		\hline
	\end{tabular}
\end{table}

\subsubsection{$XZ$ plane}
For the rays in the $XZ$ plane we set $y_i$ and $L_{yi}$ to zero 
and get
\begin{eqnarray}
\Phi_{sXZ}^{(4)} = \sum_{j=0}^3 \sum_{i=1}^f A_{j1i}x_i^{j+1},
\end{eqnarray}
where $A_{j1i}$ is the aberration coefficient $A_{j1}$ for the $i$th surface. 

The distance along an arbitrary ray between the actual and reference wavefronts at the final surface is given by $\Delta = \Phi_s/n_{1f}$. For the transverse aberration at the final image the $j$th component can be calculated as 
\begin{eqnarray}
{\rm TA}_{jXZ} = s_f \frac{\partial \Delta_j}{\partial x_f} =  \frac{s_f}{n_f}(j+1)\sum_{i=1}^f A_{j1i} x_i^j \frac{\partial x_i}{\partial x_f}
\end{eqnarray}
with $s_f\equiv s_{1f}$ and $n_f\equiv n_{1f}$. By the usual replacement of $\partial x_i/\partial x_f \to x_i/x_f$~\cite{Schroeder} we find 
\begin{eqnarray}\label{eq:TA1}
{\rm TA}_{jXZ}  =  \frac{s_f}{n_f}(j+1) A_{js}^{(f)} x_f^j,
\end{eqnarray}
where
\begin{eqnarray}
A_{js}^{(f)} = \sum_{i=1}^f A_{j1i} \left(\frac{x_i}{x_f}\right)^{j+1}.
\end{eqnarray}
The \eqref{eq:TA1} allows to calculate the transverse aberration with the marginal ray height at the last surface. This aberration can also be expressed in terms of the marginal ray height at some other surface (e.g., at the OS entrance pupil), say the $l$th one, as follows
\begin{eqnarray}\label{eq:TA2}
{\rm TA}_{jXZ}  =  \frac{s_f}{n_f}(j+1) A_{js}^{(l)} \frac{x_l^{j+1}}{x_f},
\end{eqnarray}
where
\begin{eqnarray}\label{eq:Ajs}
A_{js}^{(l)} = \sum_{i=1}^f A_{j1i} \left(\frac{x_i}{x_l}\right)^{j+1}.
\end{eqnarray}

\subsubsection{$YZ$ plane}
By setting $x_i$ and $L_{xi}$ to zero we have
\begin{eqnarray}
\Phi_{sYZ}^{(4)} = \sum_{j=0}^3 \sum_{i=1}^f \tilde B_{ji} \, y_i^{j+1}
\end{eqnarray}
with
\begin{eqnarray}
&\tilde B_{0i} = B_{02i}(L_{xi}=0),& \qquad \tilde B_{1i} = A_{12i}(L_{xi}=0) + B_{12i}, \nonumber\\
&\tilde B_{2i} = B_{22i},&  \qquad  \tilde B_{3i} = A_{32i}. \label{eq:tildeB}
\end{eqnarray}
Hence, for the $YZ$ plane the transverse aberration $j$th term at the final image is given by
\begin{eqnarray}
{\rm TA}_{jYZ}  =  \frac{s_f}{n_f}(j+1) \tilde B_{js}^{(f)} y_f^j 
\end{eqnarray}
with
\begin{eqnarray}
\tilde B_{js}^{(f)} = \sum_{i=1}^f \tilde B_{ji} \left(\frac{y_i}{y_f}\right)^{j+1},
\end{eqnarray}
which also can be rewritten in a form similar to \eqref{eq:TA2}. 
However, the numerical values of all $\tilde B$ in \eqref{eq:tildeB} are vanishing.

\subsection{Cylindrical optical system of Cassegrain type}\label{sec:cassegrain} 

The Cassegrain telescope has concave primary and convex secondary ($\mu_2>0$). 
Consider its cylindrical analog, which primary and secondary are the concave 
and convex 
cylindrical mirrors, respectively. 
Let 
$b_1=b_2=0$ and $n_{01}=n_{02}=1$. We assume the OS stop is at the primary, i.e., in front of the secondary. Then for the primary mirror with the focal length $f_1$ (obviously, for cylindrical mirror we have the focal line instead of the focal point) we take $W_1=L_{x1}=\psi_1=0$, $s_{01}=\infty$ and $s_{11}=f_1$. Using Table~\ref{tab:aberrations1}, we obtain
\begin{flalign}
A_{011}(L_{x1}=0) &= \alpha_d = 0,  \ \  A_{111}(L_{x1}=0) = \frac{\alpha_{a1}}{2} = - \frac{\theta_0^2}{2R_1}, \\
A_{211}(L_{x1}=0) &= \frac{\theta_0}{R_1},  \quad  A_{311} = -\frac{\kappa_1+1}{4R_1^3},
\end{flalign}
where $\theta_0 \equiv \theta_{01}$.

From the law of reflection with the replacement discussed below \eqref{eq:formula3} we find
\begin{eqnarray}
\psi_2=\pi-\theta_{11}=\theta_0	
\end{eqnarray}
Using Table~\ref{tab:aberrations22}, for the secondary we get
\begin{flalign}
&A_{012} = \frac{(W\theta_0)^3}{2R_2^3} \nonumber\\
&\times \left[ \kappa_2 - \left( \frac{\mu+1}{\mu-1} \right)^2 + 3R_2^2\left( \frac{1}{W} - \frac{1}{R_2} \right)^2 - R_2^2\left( \frac{1}{W} - \frac{1}{s_{02}} \right)^2 \right], \\
&A_{112} = \frac{(W\theta_0)^2}{2R_2^3} \left[ 3\kappa_2 + 3R_2^2\left( \frac{1}{W} - \frac{1}{R_2} \right)^2 - 2R_2^2\left( \frac{1}{W} - \frac{1}{s_{02}} \right)^2 \right], \\
&A_{212} = \frac{W\theta_0}{R_2^2} \left[ \frac{\kappa_2}{R_2} + \left( \frac{1}{W} - \frac{1}{R_2} \right) \frac{\mu+1}{\mu-1} \right], \\
&A_{312} = \frac{1}{4R_2^3} \left[ \kappa_2 + \left( \frac{\mu+1}{\mu-1} \right)^2 \right],
\end{flalign}
where $W\equiv W_2$, $\mu\equiv\mu_2$ and the overall signs are taken positive since the mirror is convex. From \eqref{eq:Ajs} for the marginal ray height at the primary $x_l=x_1$ we have
\begin{eqnarray}
A_{js}^{(1)} = A_{j11} + A_{j12} \left( \frac{x_2}{x_1} \right)^{j+1}
\end{eqnarray}
with $j=0,1,2,3$. In terms of the variables $k_x=x_2/x_1$ and $\rho=R_2/R_1$ we obtain
\begin{eqnarray}
A_{3s}^{(1)} =  -\, \frac{1}{4R_1^3} \,F_{XZ}(\kappa_1,\kappa_2,\mu,\rho,k_x),
\end{eqnarray}
where 
\begin{eqnarray}
F_{XZ}(\kappa_1,\kappa_2,\mu,\rho,k_x) = \kappa_1 + 1 - \frac{k_x^4}{\rho^3} \left[ \kappa_2 + \left( \frac{\mu+1}{\mu-1} \right)^2 \right]
\end{eqnarray}
with positive $\rho$. Finally, using the relation of $s_{12}=k_xf$, where $f$ is the OS focal length, the transverse spherical aberration can be written as
\begin{eqnarray}
{\rm TA}_{3XZ}  =  4f A_{3s}^{(1)} x_1^3 = - f\left(\frac{x_1}{R_1}\right)^3 F_{XZ}(\kappa_1,\kappa_2,\mu,\rho,k_x). 
\end{eqnarray}
From the relation of $F_1=R_1/(4x_1)$, where $F_1$ is the primary mirror focal ratio, we get
\begin{eqnarray}
{\rm TA}_{3XZ}  =  -\, \frac{f}{64F_1^3} F_{XZ}(\kappa_1,\kappa_2,\mu,\rho,k_x),
\end{eqnarray}
which coincides with the result for a rotationally symmetric OS~\cite{Schroeder} as expected.

Expressions for the rest of the transverse aberrations for the rays in the $XZ$ and $YZ$ planes can be determined using the discussed procedure. 
Condition $F_{XZ}=0$ for zero elliptic aberration in the $XZ$ plane coincides with the known condition of the spherical aberration compensation for the rotationally symmetric Cassegrain telescope. One of the possibilities for such compensation provides the classical Cassegrain telescope with the paraboloidal primary and hyperboloidal secondary, i.e.: 
\begin{eqnarray}
\kappa_1=-1, \qquad   \kappa_2=-\left(\frac{\mu+1}{\mu-1}\right)^2.
\end{eqnarray}
If the discussed OS is embedded to an anamorphic OS, which optical axis is the $Z$ axis, then the function $F_{XZ}$ 
enters respective equation for the full system leading aberrations compensation.

\section{Conclusion}

In this paper the method of calculation of monochromatic aberrations for the optical systems with the symmetry of a generic cylinder have been developed. In particular, the aberration coefficients for calculations of the aberrations for an arbitrary paraxial ray were obtained in certain approximations. Interestingly, the second coma is vanishing together with the distortion for a single mirror surface ($\alpha_{c2}=0=\alpha_d$ in Table~\ref{tab:aberrations1}). Further, the derived analytical results have been applied to an example optical system, namely, the cylindrical analog of the Cassegrain telescope.

We notice that choice of suitable approximations, in general, depends on the parameters of the optical system and requires further investigation. 

We remark that the discussed optical systems with translational invariance and single-plane symmetry can serve as components of more advanced anamorphic optical systems, in particular, narrow field-of-view anamorphic telescopes, 
which is a subject of the author's patent application Nr.\,P.444521 submitted to the Patent Office of the Republic of Poland on April 21, 2023. 

\begin{backmatter}

\bmsection{Disclosures} The author declares no conflicts of interest.

\bmsection{Data availability} No data were generated or analyzed in the presented research.

\end{backmatter}





\end{document}